\documentclass[letterpaper, 10 pt, conference]{ieeeconf}  
\IEEEoverridecommandlockouts

\newcommand{\revised}{black}
\usepackage{amsmath,amssymb,amsfonts}
\newtheorem{proposition}{Proposition}
\usepackage{algorithmic}
\usepackage{graphicx}
\usepackage{subcaption}
\usepackage{textcomp}
\usepackage{cite}
\usepackage{xcolor}

\newtheorem{remark}{Remark}
\usepackage{comment}

\newif\ifjournal
\journaltrue
\newif\ifpreprint
\preprinttrue \journalfalse 

\begin{document}

\title{\LARGE \bf
Feedback Control of a Recirculating Bioreactor with Electrophoretic Removal of Inhibitory Extracellular DNA
}

\author{Antonio Spallone$^{1,2,3}$, Davide Fiore$^{4}$, Fabrizio Cartenì$^{3}$, and Mario di Bernardo$^{1,5}$%
%
\thanks{$^{1}$Department of Electrical Engineering and Information Technology, University of Naples Federico II, Naples, Italy.
        {\tt\small \{antonio.spallone, mario.dibernardo\}@unina.it}}%
\thanks{$^{2}$Department of Electrical and Information Engineering, Polytechnic of Bari, Bari, Italy}%
\thanks{$^{3}$ Department of Agricultural Sciences, University of Naples Federico II, Naples, Italy. {\tt\small fabrizio.carteni@unina.it}}%
\thanks{$^{4}$Department of Mathematics and Applications “R. Caccioppoli,”
        University of Naples Federico II, Naples, Italy.
        {\tt\small davide.fiore@unina.it}}%
\thanks{$^{5}$Scuola Superiore Meridionale, Naples, Italy.}%
}

\pagestyle{empty}

\maketitle
\thispagestyle{empty}

\begin{abstract}
Extracellular DNA accumulation in recirculating bioprocesses inhibits microbial growth and reduces productivity. We consider a continuous bioreactor with a recirculating loop and an electrophoretic filtration unit for selective DNA removal, and develop a feedback control framework combining online state and parameter estimation via an Unscented Kalman Filter with an adaptive Model Predictive Controller that jointly optimizes dilution rate and filtration activation.
Closed-loop simulations under nominal and perturbed conditions show that the 
{\color{\revised} addition of the filtration unit enables the proposed control strategy to} achieve significantly higher cumulative profit while keeping DNA concentration below the inhibition threshold.
\end{abstract}

\section{Introduction}
\label{sec:introduction}

Continuous bioprocesses based on chemostat operation are widely used in biomanufacturing because they enable near steady-state operation, stable productivity, and consistent product quality \cite{Meena2021,JBiotec2015}. 
{\color{\revised}The control of such systems has been extensively studied and various control strategies, like model predictive control, adaptive control, and observer-based strategies, have been applied to fed-batch, continuous, and perfusion processes for feed-rate regulation, productivity optimization, and robust operation under parameter uncertainty \cite{BarzagaMartell2021,Pappenreiter2022,Pataro2023}.}
However, conventional chemostats require a continuous supply of fresh medium and a continuous discharge of culture broth, therefore, in view of increasingly stringent environmental requirements and circular-economy targets, reduced-discharge and recirculating configurations are becoming increasingly relevant \cite{Fernandes2023,Klok2025}.

A critical drawback of recirculation is the possible accumulation of inhibitory compounds in the culture medium that may reduce growth and productivity, especially when the inhibitory species are not removed together with the outlet stream \cite{Hsiao1994,Borole2009}.
{\color{\revised}Several works have addressed this problem. Examples include acetate accumulation in \textit{E. coli} cultures, ethanol inhibition in continuous fermentation, and product-inhibited fermentations coupled with in-situ removal units \cite{Abadli2022,Skupin2022}.}
Furthermore, extracellular DNA has recently been identified as a relevant inhibitor of cell proliferation when its concentration exceeds a critical threshold \cite{deAlteriis2023,Mazzoleni2015a}. 

{\color{\revised}Motivated by this issue, in this letter we consider a recirculating bioreactor equipped with an electrophoretic filtration module, currently under experimental validation, which exploits the electric charge of DNA molecules to selectively remove extracellular DNA from the recirculating stream. The filtration concept is covered by a patent held by NOSELF s.r.l.~\cite{Mazzoleni2014p}; a peer-reviewed characterisation of the prototype is currently in preparation. 
Contrary to the solutions proposed in the literature, in which inhibition is mitigated indirectly by adjusting the feed rate, dilution rate, perfusion rate, or operating set-point, while the inhibitory compound remains an internal process variable, here the inhibitory compound is explicitly modeled as a dynamic state and its removal is treated as a control action. This leads to a hybrid economic control problem in which a continuous dilution-rate input must be coordinated with a binary filtration-activation input.
The contribution of this letter is therefore application-driven but, to the best of our knowledge, addresses a control problem that has not been previously formulated for recirculating bioreactors with extracellular-DNA inhibition. 
Specifically, we develop a nonlinear model of the process, analyze the practical observability and identifiability of the augmented state-parameter system, and combine an Unscented Kalman Filter with a receding-horizon controller. The UKF is used to estimate both the unmeasured extracellular-DNA concentration and uncertain kinetic parameters, while the MPC optimizes the dilution rate and the activation of the electrophoretic filtration unit to maximize an economic gain.
The performance and robustness of the closed-loop system are assessed through extensive simulations which highlight a significant increase of the economic gain when the filtration unit is employed. 
}

\section{Bioreactor Model and Structural Observability}
\label{sec:model}

\begin{figure}
    \centering
    \includegraphics[width=0.76\linewidth]{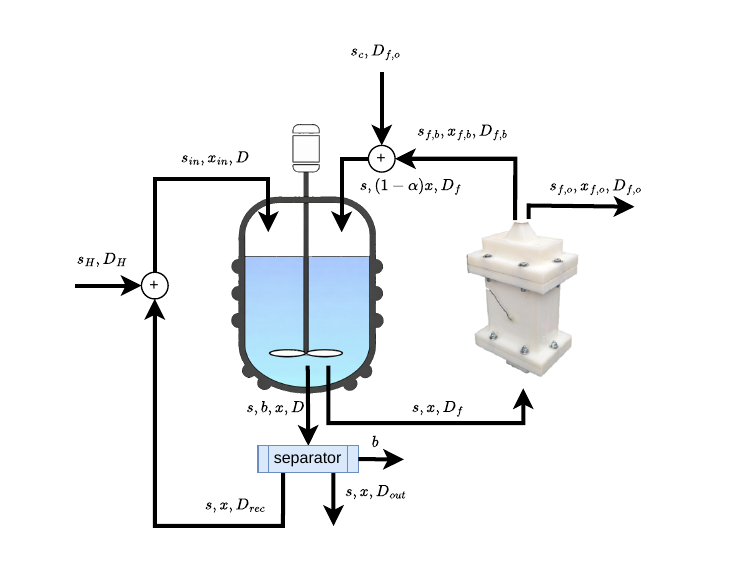}
    \caption{\footnotesize Schematic representation of the bioreactor with a recirculating loop (on the left) and an electrophoretic filtration unit (on the right).}
    \label{fig:schema}
\end{figure}
We consider a perfectly mixed chemostat bioreactor with a recirculation loop and an electrophoretic filtration unit (Fig.~\ref{fig:schema}). The mathematical model describes the growth of a microbial population inside the reactor \cite{SmithWaltman1995,Monod1949}, with additional mechanisms accounting for extracellular DNA accumulation and its removal through the filtration unit. The system dynamics are given by
\begin{equation}
\begin{aligned}
\dot b &= \mu(s,x)b - D(t)b,\\
\dot s &= -\frac{1}{Y}\mu(s,x)b + D(t)(s_\mathrm{in}-s),\\
\dot x &= c\,\mu(s,x)b - (D(t) -D_\mathrm{rec}(t))x - \delta(t)D_{f}\alpha x  ,
\end{aligned}
\label{eq:model}
\end{equation}
where $b(t) \in \mathbb{R}_{\geq 0}$ is the biomass concentration $[gL^{-1}]$, $s(t) \in [0, s_{\mathrm{in}}]$ is the substrate concentration $[gL^{-1}]$, and $x(t) \in \mathbb{R}_{\geq 0}$ is the concentration of extracellular DNA released during growth $[ng \mu L^{-1}]$. The constant $s_{\mathrm{in}}$ denotes the substrate concentration in the inlet flow, the parameter $Y > 0$ is the biomass yield coefficient, and $c > 0$ is the rate of extracellular DNA production per unit biomass growth.

The function $\mu(s,x)$ is the specific growth rate of the microbial population, assumed here to be \textit{Saccharomyces cerevisiae}, which, in the absence of inhibition, follows standard Monod kinetics. To account for the inhibitory effect of extracellular DNA accumulation, the growth rate is defined as
\begin{equation}
\mu(s,x) =
\mu_{\max}
\frac{s}{K_s+s}
\left(1-\frac{x}{x_{\mathrm{crit}}}\right),
\label{eq:growth_rate}
\end{equation}
where $\mu_{\max} > 0$ is the maximum specific growth rate, $K_s > 0$ is the Monod half-saturation constant, and $x_{\mathrm{crit}} > 0$ is the critical DNA concentration at which growth is completely inhibited.  

The reactor operates with partial recirculation of the outlet stream, hence the total dilution rate $D(t)$ is decomposed as $D(t)=D_\mathrm{rec}(t)+D_\mathrm{H}(t)$,
where $D_{\mathrm{rec}}(t)$ is the recirculated fraction and $D_\mathrm{H}(t)$ is a concentrated make-up stream with substrate concentration $s_\mathrm{H}$. The make-up stream compensates for nutrient depletion in the recirculated liquid so that the effective inlet concentration equals $s_{\mathrm{in}}$. This mass-balance constraint yields the constraint $D_\mathrm{rec}(t)s(t) + D_\mathrm{H}(t)s_H = D(t)s_\mathrm{in}$.
Note that the previous equation becomes singular when $s(t) = s_\mathrm{H}$. In the operating regime considered here, $s_\mathrm{H} \gg s_{\mathrm{in}}$ so that $s(t) < s_{\mathrm{in}} < s_\mathrm{H}$ holds throughout, and the singularity is never reached. {\color{\revised}Moreover, under these assumptions $D_{\mathrm{rec}} \le D$, and both $D_\mathrm{rec}$ and $D_\mathrm{H}$ are fully determined by $D(t)$ and the current substrate concentration $s(t)$, so that the controller has a single degree of freedom in the continuous input, namely $D(t)$.}

Extracellular DNA is selectively removed by the electrophoretic filtration unit installed in the recirculation loop. The binary variable $\delta(t) \in \{0,1\}$ models filter activation: when $\delta(t) = 1$, DNA is removed at rate $D_{f} \alpha\, x$, where $D_{f}$ is the filtration flow rate and $\alpha \in (0,1]$ is the filtration efficiency. The control input vector is therefore $u(t)=[ D(t) \ \delta(t)]^\top$, subject to the constraints $0 \le D(t) \le D_{\max}$ and $\delta(t)\in\{0,1\}$.

\subsection{Model calibration}

\begin{figure}[t]
\centering

\begin{subfigure}{0.32\linewidth}
\centering
\begin{tabular}{c c}

\includegraphics[width=\linewidth]{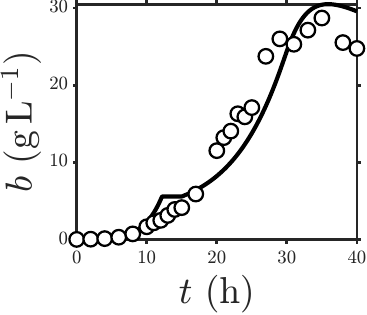} \\

\end{tabular}
\end{subfigure}
\hfill
\begin{subfigure}{0.32\linewidth}
\centering
\begin{tabular}{c c}
\includegraphics[width=\linewidth]{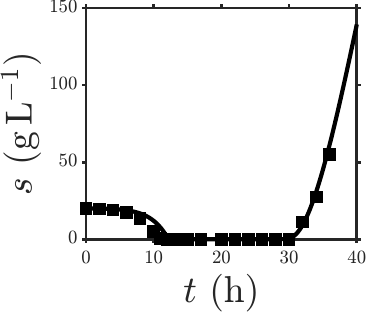} \\
\end{tabular}
\end{subfigure}
\hfill
\begin{subfigure}{0.32\linewidth}
\centering
\begin{tabular}{c c}
\includegraphics[width=\linewidth]{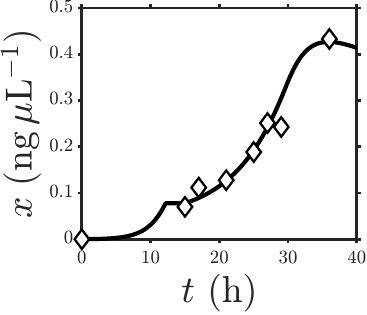} \\
\end{tabular}
\end{subfigure}

\caption{\footnotesize Model fit against experimental data for biomass $b$, substrate $s$, and extracellular DNA $x$.}
\label{fig:fitting}

\end{figure}

The kinetic parameters were identified from experimental datasets describing aerobic growth of \textit{Saccharomyces cerevisiae} in fed-batch conditions, reported in~\cite{deAlteriis2023, Mazzoleni2015}. Since the available data correspond to fed-batch operation, the model~\eqref{eq:model} was reduced to its classical chemostat formulation by setting $D_{\mathrm{rec}} = 0$ and $\delta = 0$. Parameter estimation was performed by nonlinear least squares, minimizing the discrepancy between model predictions and experimental measurements of biomass, substrate, and extracellular DNA concentrations. The resulting parameter values and their confidence intervals are summarized in Table~\ref{tab:parameters}, and the fitting results are shown in Fig.~\ref{fig:fitting}.
The inhibition threshold $x_{\mathrm{crit}}$ was obtained from prior experiments on extracellular DNA effects in microbial systems, while the filtration efficiency $\alpha$ was estimated from preliminary tests on a prototype of the electrophoretic filter.

\begin{table}[!b]
\centering
\caption{\footnotesize Model parameters and their assigned values. }
\label{tab:parameters}
\begin{tabular}{lcccc}
\hline
Parameter  & Value & CI & Units \\
\hline
$\mu_{\max}$  & 0.466 & $\pm 3.74\%$ & h$^{-1}$ \\
$K_s$  & 0.02285 & $\pm 1.25\%$ & g\,L$^{-1}$ \\
$c$  & 0.01404 & $\pm 2.93\%$ & $\mathrm{g}\,\mathrm{L}^{-1}$ \\
$Y$  & 0.2779 & $\pm 0.52\%$ & --\\
$\alpha$  & 0.72 & -- & -- \\
$D_{\max}$  & 0.6 & -- & h$^{-1}$\\
$D_{\min}$  & 0.05 & -- & h$^{-1}$\\
$x_{\text{crit}}$  & 0.48 & -- & ng\,$\mu$L$^{-1}$ \\
$D_f$  & 0.4 & -- & h$^{-1}$ \\
$s_\mathrm{in}$ & 20 & -- & g\,$\mathrm{L}^{-1}$\\
$s_H$ & 200 & -- & g\,$\mathrm{L}^{-1}$\\
$\lambda$ & 2.4 & -- & --\\
\hline
\end{tabular}
\end{table}

\subsection{Structural identifiability and observability}
\label{sec:observability}

{\color{\revised}Reliable model-based control requires that the relevant system states and uncertain kinetic parameters can be reconstructed from the available measurements. We therefore analyze the structural observability and parameter identifiability properties of the nonlinear bioreactor model~\eqref{eq:model}. The aim of this analysis is not to establish global observability over the whole state space, but rather to verify local reconstructability in the admissible operating region where the bioreactor is actually run.}

Following the standard augmented-state formulation commonly adopted in joint state and parameter estimation problems~\cite{Kandepu2008UKF,GoveHollinger2006DualUKF}, we augment the system state with the uncertain kinetic parameters:
\begin{equation}
z = [b,s,x,\theta]^\top,
\qquad
\theta = [\mu_{\max},K_s,c,Y]^\top .
\label{eq:augmented_state}
\end{equation}
The inhibition threshold $x_{\mathrm{crit}}$ and the filtration efficiency $\alpha$ are assumed known and constant. The augmented dynamics can be written as
\begin{equation}
\dot z(t)=
\begin{bmatrix}
f(\xi(t),u(t),\theta)+w_\xi(t)\\
w_\theta(t)
\end{bmatrix},
\label{eq:augmented_system}
\end{equation}
where $\xi=[b\;s\;x]^\top$, $w_\xi(t)$ accounts for modelling uncertainties, and $w_\theta(t)$ models slow parameter variations and unmodelled dynamics. The structural observability analysis below is carried out on the deterministic part of~\eqref{eq:augmented_system}, i.e., by setting the uncertainty terms to zero.

Measurements are collected at discrete sampling instants $t_k$. The measured output is $y_k = H_k\,\xi_k + v_k$, 
where $v_k$ denotes measurement noise. In the sensing configuration considered here, biomass and substrate concentrations are available at every sampling instant, whereas extracellular DNA measurements are only intermittently available. The measurement matrix is therefore
\begin{equation}
 H_k =
\begin{bmatrix}
 1 & 0 & 0 \\
 0 & 1 & 0
 \end{bmatrix},
 k \notin \mathcal{K}_x,\
\mbox{or}\
\begin{bmatrix}
 1 & 0 & 0 \\
 0 & 1 & 0 \\
 0 & 0 & 1
 \end{bmatrix},
 \ k \in \mathcal{K}_x,
\end{equation}
where $\mathcal{K}_x$ denotes the set of sampling instants at which extracellular DNA measurements are available.

The possibility of reconstructing the hidden state $x$ and the uncertain parameter vector $\theta$ depends on the observability properties of the augmented nonlinear system~\eqref{eq:augmented_system}. To assess local weak observability and parameter identifiability, we employ the nonlinear observability rank condition introduced by Hermann and Krener~\cite{HermannKrener1977}. Let $\mathcal{O}(z)$ denote the observability matrix generated by the gradients of the output map and its successive Lie derivatives along the augmented dynamics.
{\color{\revised}In nonlinear bioprocess models, global observability over arbitrary state and parameter values is generally neither expected nor required for control purposes. What is relevant here is observability in the biologically meaningful operating region characterized by positive biomass and substrate concentrations, extracellular DNA below the complete-inhibition threshold, and kinetic parameters within admissible ranges. This type of local structural analysis is commonly used in biological and bioprocess models, where identifiability and observability are assessed in the neighbourhood of feasible trajectories or within parameter regions of practical interest~\cite{Czyniewski2022}.}
We therefore consider the most challenging sensing configuration, in which extracellular DNA measurements are unavailable for all sampling instants, i.e., $k \notin \mathcal{K}_x$ for all $k$. In this case, extracellular DNA is treated as a completely unmeasured state and only biomass and substrate concentrations are available. The rank condition $\mathrm{rank}\big(\mathcal{O}(z^*)\big)=\mathrm{dim}(z)$

is then evaluated over the admissible operating region. {\color{\revised}Since the high-order Lie derivatives of this rational nonlinear model rapidly become algebraically cumbersome, the analysis combines analytical derivations of the first independent codirections with a numerical rank verification of the complete observability matrix, as commonly done in nonlinear observability studies~\cite{Shi2022}. The result should therefore be interpreted as a local practical observability and identifiability property over the considered operating region.}

The preceding analysis motivates the following result.

\begin{proposition}[{\color{\revised}Practical} observability and identifiability]
\label{prop:obs}
Consider the augmented system~\eqref{eq:augmented_system} with output $y_k = H_k \xi_k$, where only $b$ and~$s$ are measured. Suppose that:
\begin{enumerate}
    \item \textcolor{\revised}{the dilution rate satisfies $D(t) > 0$ for all $t \geq 0$;}
    \item the concentrations satisfy $b(t) > 0$, $s(t) > 0$, and $x(t)<x_{\mathrm{crit}}$ for all $t \geq 0$.
\end{enumerate}
Then the augmented system satisfies the Hermann--Krener local observability rank condition throughout the admissible operating region considered in this work. Consequently, the extracellular DNA concentration $x$ and the uncertain parameter vector~$\theta$ are locally identifiable from biomass and substrate measurements within this region.
\end{proposition}

\ifpreprint
The proof of Proposition~\ref{prop:obs} is reported in Appendix~\ref{app:proof_prop1}.
\fi

\textcolor{\revised}{As an additional  verification of the above observability-identifiability argument, the augmented model was analysed with the symbolic MATLAB toolbox STRIKE-GOLDD \cite{DazSeoane2022}. The analysis returned full input-state-parameter observability, confirming that all states are observable and all augmented parameters are locally structurally identifiable under the considered output configuration. The informativeness of the closed-loop inputs for the joint UKF estimator is then assessed a posteriori through the bounded condition number of the augmented covariance matrix $P_{z,k}$ along the Monte Carlo trajectories (see Section~\ref{sec:numerical_validation}).
}

\section{MPC Formulation}
\label{sec:mpc}

We first consider the deterministic case in which the state $\xi=[b\;s\;x]^\top$ and the parameter vector $\theta=[\mu_{\max},K_s,c,Y]^\top$ are known exactly, while the practical case with online estimation is addressed later in Section~\ref{sec:practical_implementation}.
At each sampling instant $k$, the MPC predicts the evolution of~\eqref{eq:model} from the current state $\bar{\xi}_{k|k}=\xi_k$ over a horizon of $N$ steps as
\begin{equation}
\label{eq:prediction}
\bar{\xi}_{k+i+1|k} = F(\bar{\xi}_{k+i|k},\,u_{k+i|k},\,\theta),
\quad i=0,\dots,N-1,
\end{equation}
where $F$ denotes the numerical integration of~\eqref{eq:model} over
one sampling interval $T_s$, and $u_{k+i|k} = [D_{k+i|k} \, \delta_{k+i|k}]^\top $, 
with $D_{k+i|k}$ the dilution rate and $\delta_{k+i|k}\in\{0,1\}$ the filtration-activation command.

The stage cost is chosen to directly reflect the economic objective of the process, yielding the following economic MPC problem~\cite{Angeli2012,Rawlings2012}:
\begin{equation}
\label{eq:mpc_cost}
\max_{u_{k|k},\dots,u_{k+N-1|k}}
\sum_{i=0}^{N-1}
\Bigl[
D_{k+i|k}\,\bar{b}_{k+i|k}
-
\lambda\,\delta_{k+i|k}\,D_f
\Bigr]T_s
\end{equation}
subject to~\eqref{eq:prediction} and, for $i=0,\dots,N-1$,
\begin{align}
D_{\min} &\le D_{k+i|k} \le D_{\max}, \label{eq:mpc_D}\\
\delta_{k+i|k} &\in \{0,1\}, \label{eq:mpc_delta}\\
\bar{b}_{k+i|k} &\ge b_{\min}, \label{eq:mpc_bmin}\\
0 \;\le\; &\bar{x}_{k+i|k} \le x_{\max}. \label{eq:mpc_xmax}
\end{align}
The term $D\bar{b}$ in~\eqref{eq:mpc_cost} represents the instantaneous harvested biomass productivity, while $\lambda\,\delta\,D_f$ penalizes the energy cost of activating the filtration unit. The cost-to-price ratio $\lambda = c_f/p_b$ normalizes the filtration operating cost $c_f$ by the biomass market value $p_b$, and has been set to $\lambda=2.4$ based on industrial yeast market data and preliminary energy-consumption estimates of the filtration module.

{\color{\revised}
The constraint on~$\bar{b}$ prevents the MPC from predicting trajectories in which the microbial population is lost, whereas the upper bound~$x_{\max}$ on the predicted extracellular-DNA concentration keeps the operating point within the admissible region defined in Proposition~\ref{prop:obs}.
The upper bound $x_{\max}$ in constraint~\eqref{eq:mpc_xmax} is chosen as a conservative safety margin below the complete-inhibition threshold $x_{\mathrm{crit}}$. Specifically, $x_{\max}$ is selected as the largest DNA concentration at which the growth rate at substrate saturation ($s \approx s_{\mathrm{in}}$) balances the minimum dilution rate $D_{\min}$, that is, $\mu(s_\mathrm{in}, x_{\max})=D_{\min}$,
which gives
\begin{equation}
\label{eq:xmax_condition}
\tfrac{x_{\max}}{x_{\mathrm{crit}}}
= 1 -
\tfrac{D_{\min}(K_s+s_{\mathrm{in}})}
{\mu_{\max}\,s_{\mathrm{in}}}.
\end{equation}
This choice ensures that, within the MPC prediction horizon, possible high extracellular-DNA concentration does not prevent the biomass to recover to safe concentration levels.

}

\ifpreprint
\begin{remark}
No terminal cost or terminal constraint is included in~\eqref{eq:mpc_cost}--\eqref{eq:mpc_xmax}. In the economic MPC literature, closed-loop stability and performance guarantees for terminal-free formulations can be established under dissipativity and turnpike conditions~\cite{Grune2013,GruneStieler2014}; verifying these properties analytically for the present nonlinear system is nontrivial and is left for future work. In the simulations reported in Section~\ref{sec:validation_det}, the prediction horizon $N$ is chosen longer than the dominant process transient, and closed-loop constraint satisfaction and economic performance are assessed empirically across all Monte Carlo scenarios.

The presence of the binary variable $\delta_{k+i|k}$ makes problem~\eqref{eq:mpc_cost}--\eqref{eq:mpc_xmax} a mixed-integer nonlinear programming (MINLP). In the closed-loop simulations, MPC feasibility was monitored at every sampling instant; the results reported in Table~\ref{tab:mc_feasibility} show that a finite admissible solution was returned in $100\%$ of optimization calls across all scenarios.

The MPC cost~\eqref{eq:mpc_cost} does not include an explicit penalty on switching of the binary filtration variable $\delta$.
This is justified by the time-scale separation between the filtration dynamics and the MPC sampling interval: the characteristic response time of the electrophoretic filtration unit is significantly shorter than $T_s$, so that transients associated with filter activation and deactivation are negligible over one sampling step.
\begin{table}[t]
\centering
\caption{Closed-loop feasibility and constraint satisfaction
over the $50$ Monte Carlo scenarios.}
\label{tab:mc_feasibility}
\begin{tabular}{lcc}
\hline
Metric & With filtration & Baseline \\
\hline
Scenarios                        & $50$                & $50$                \\
Finite-objective MPC calls [\%]  & $100$               & $100$               \\
Runs with $b_k < b_{\min}$       & $0$                 & $0$                 \\
Runs with $x_k > x_{\max}$       & $0$                 & $18$                \\
Max violation of $x_{\max}$      & $0$                 & $3.07\times10^{-2}$ \\
Max $x_k/x_{\max}$               & $0.494$             & $1.073$             \\
\hline
\end{tabular}
\end{table}
\end{remark}

\fi

\subsection{Closed-loop validation}
\label{sec:validation_det}
{\color{\revised}Next, we compare the closed-loop performance of the proposed MPC with electrophoretic filtration ($\delta_k$ optimized) to those of a ``baseline'' MPC in which filtration was disabled ($\delta_k=0$ for all $k$). Both controllers were initialized at $b(0)=0.1\,\mathrm{g\,L^{-1}}$, $s(0)=20\,\mathrm{g\,L^{-1}}$, $x(0)=0\,\mathrm{ng\,\mu L^{-1}}$.
Specifically, we considered a paired Monte Carlo
study comprising $50$ scenarios in which the kinetic parameters $\theta = [\mu_{\max}, K_s, c, Y]^\top$ have been drawn uniformly within the admissible ranges in Table~\ref{tab:parameters}, and compared the performance of the two controllers by evaluating four metrics:

(i) the cumulative economic return at step $i$ was computed as
\begin{equation}
\label{eq:Jcl_validation}
J_i=\sum_{k=0}^{i-1}
\bigl(D_k\,b_k - \lambda\,\delta_k\,D_f\bigr)\,T_s,
\qquad i=1,\dots,N_{\mathrm{sim}},
\end{equation}
(ii) the value of the biomass at the end of the simulation, (iii) the maximum
extracellular-DNA concentration, i.e. $x_{\max}^{\mathrm{cl}}=\max_{0\le k\le N_{\mathrm{sim}}}x_k$, 
and (iv) the mean value of $D$.


The improvement of the filtration-based MPC controller with respect to the baseline has been quantified using two-sided paired $t$-tests on the within-scenario relative differences, defined as
\begin{equation}
\label{eq:paired_difference}
\Delta m_j^{(\%)} =
100\,
\tfrac{
m_j^{\mathrm{filter}} - m_j^{\mathrm{baseline}}
}{
m_j^{\mathrm{baseline}}
},
\qquad j=1,\dots,50,
\end{equation}
whose values mediated over all simulations are in Tab.~\ref{tab:mc_statistical_comparison}.}

\subsubsection{Closed-loop control performance}
The filtration-enabled MPC achieved a statistically significant improvement in terminal cumulative economic return ($J_{N_{sim}}$), with an average relative increase of $51.35\%$ over the simulated horizon ($p \approx 0$). This improvement was not associated with a significant change in the final biomass concentration ($0.26\%$, $p=0.756$), indicating that the higher economic return was mainly due to a more efficient closed-loop operating policy rather than to a higher terminal biomass. At the same time, active filtration reduced the maximum extracellular DNA concentration by $56.64\%$ on average ($p\approx 0$), while allowing the controller to sustain a higher mean dilution rate, increased by $90.67\%$ ($p \approx 0$). Overall, these results show that the filtration-enabled MPC better exploits the admissible operating region, consistently with the economic objective~\eqref{eq:Jcl_validation}.
{\color{\revised}Moreover, the filtration-enabled MPC satisfied all state constraints in every scenario, whereas the baseline exceeded the DNA bound $x_{\max}$ in \eqref{eq:xmax_condition} in $18$ out of $50$ scenarios.}

\begin{table*}[!t]
\centering
\caption{\footnotesize {\color{\revised}Comparison between the filtration-enabled MPC and the baseline. Main values refer to the UKF-based Monte Carlo analysis of Section~\ref{sec:numerical_validation}, while parentheses report the perfect-state-feedback validation of Section~\ref{sec:validation_det}. Relative changes are computed with respect to the corresponding baseline, and $p$-values come from two-sided paired $t$-tests on the within-scenario relative differences~\eqref{eq:paired_difference}.}}
\label{tab:mc_statistical_comparison}
\scriptsize
\setlength{\tabcolsep}{3pt}
\renewcommand{\arraystretch}{1.05}
\begin{tabular}{lcccc}
\hline
Metric & With filtration & Baseline & Relative change [\%] & $p$-value \\
\hline

$J_{N_{\mathrm{sim}}}$
& $53.91\!\pm\!6.81$ {\scriptsize($47.14\!\pm\!6.64$)}
& $32.24\!\pm\!2.77$ {\scriptsize($31.55\!\pm\!5.56$)}
& $67.02\!\pm\!14.96$ {\scriptsize($51.35\!\pm\!19.76$)}
& $\approx 0$ \\

Final $b$ [g\,L$^{-1}$]
& $5.399\!\pm\!0.428$ {\scriptsize($4.77\!\pm\!0.58$)}
& $5.401\!\pm\!0.206$ {\scriptsize($4.76\!\pm\!0.50$)}
& $-0.13\!\pm\!6.00$ {\scriptsize($0.26\!\pm\!5.85$)}
& $0.943$ {\scriptsize($0.756$)} \\

Max $x$ [ng\,$\mu$L$^{-1}$]
& $0.176\!\pm\!0.013$ {\scriptsize($0.175\!\pm\!0.018$)}
& $0.427\!\pm\!0.021$ {\scriptsize($0.403\!\pm\!0.032$)}
& $-58.79\!\pm\!2.45$ {\scriptsize($-56.64\!\pm\!2.75$)}
& $\approx 0$ \\

Mean $D$ [h$^{-1}$]
& $0.263\!\pm\!0.030$ {\scriptsize($0.264\!\pm\!0.021$)}
& $0.126\!\pm\!0.013$ {\scriptsize($0.142\!\pm\!0.025$)}
& $108.49\!\pm\!20.17$ {\scriptsize($90.67\!\pm\!27.98$)}
& $\approx 0$ \\

\hline
\end{tabular}
\end{table*}

\subsubsection{Computational tractability}
{\color{\revised}To assess real-time implementability, the MPC solution time was recorded at each optimization call. 
The MINLP was solved at each sampling instant using the Genetic Algorithm solver available in MATLAB R2024b on 13th Gen Intel Core i9-13900HX processor. The prediction horizon was set to $N=18$, with a sampling time of $T_s = 0.75\,\mathrm{h}$ and a simulation length of $N_{\mathrm{sim}} = 67$. 
All optimization calls completed within the sampling interval $T_s=2700\,\mathrm{s}$. The worst-case solution time of $35.36\,\mathrm{s}$ corresponds to $1.31\%$ of $T_s$, confirming computational tractability at the considered sampling rate.}
\ifpreprint
Complete solver settings and MPC computation-time statistics are reported in Table~\ref{tab:solver_settings} and Table~\ref{tab:mpc_times_ukf}.
\begin{table}[t]
\centering
\caption{MPC solver and simulation settings.}
\label{tab:solver_settings}
\begin{tabular}{ll}
\hline
Setting & Value \\
\hline
Solver & Genetic algorithm \\
Processor & 13th Gen Intel Core i9-13900HX \\
$N$ & $18$ \\
$T_s$ & $0.75\,\mathrm{h}$ \\
$N_{\mathrm{sim}}$ & $67$ \\
Maximum generations & $600$ \\
Function tolerance & $10^{-4}$ \\
Initial population & User-defined \\
Parallel evaluation & Disabled \\
\hline
\end{tabular}
\end{table}
\begin{table}[t]
\centering
\caption{MPC solution times over all closed-loop calls in the Monte Carlo scenarios. Percentages are with respect to $T_s=2700\,\mathrm{s}$.}
\label{tab:mpc_times_ukf}
\begin{tabular}{lcccc}
\hline
 & \multicolumn{2}{c}{With filtration}\\
\cline{2-3}\cline{4-5}
 & Time [s] & $T_s\,[\%]$\\
\hline
Mean        & $12.40$ & $0.46$ \\
$95$th pct. & $21.56$ & $0.80$ \\
Worst case  & $36.69$ & $1.36$ \\
\hline
\end{tabular}
\end{table}
\fi


\section{Towards Practical Implementation}
\label{sec:practical_implementation}

The control design in  Section~\ref{sec:mpc} assumes exact knowledge of the state $\xi$ and the parameter vector $\theta$, but in practice, they are not directly accessible.
{\color{\revised}Biomass and substrate concentrations are measurable online, but extracellular DNA is only available at sparse intervals, and the kinetic parameters are uncertain and subject to biological variability.} To address this, we coupled the  MPC with an Unscented Kalman Filter
(UKF)~\cite{Julier1997,Wan} that recursively estimates the full augmented state $z=[b,s,x,\theta]^\top$ defined in \eqref{eq:augmented_state}.

\subsection{Adaptive UKF-MPC architecture}
\label{sec:ukf_mpc}

{\color{\revised}The augmented state is propagated according to the model
\begin{equation}
\label{eq:ukf_augmented_model}
\xi_{j+1} = F_{\mathrm{UKF}}(\xi_j,u_j,\theta_j) + w_{\xi,j},
\quad
\theta_{j+1} = \theta_j + w_{\theta,j},
\end{equation}
where $F_{\mathrm{UKF}}$ denotes the numerical integration of~\eqref{eq:model} over the UKF update interval $T_{\mathrm{UKF}}=2\,\mathrm{min}$, and $w_{\xi,j}$, $w_{\theta,j}$ are white Gaussian noises with covariances $Q_\xi$ and $Q_\theta$, respectively. The random-walk parametrization of $\theta$ is standard in joint state-parameter estimation for bioprocesses \cite{Kandepu2008UKF,GoveHollinger2006DualUKF} and allows the filter to track slow parameter variations due to biological variability or model mismatch.}

{\color{\revised}The measurement model follows~\eqref{eq:augmented_system}, with biomass and substrate measurements available at every UKF update step, whereas extracellular DNA measurements are available only every $6\,\mathrm{h}$, i.e., at sparse instants $k\in\mathcal{K}_x$. The observability analysis of Section~\ref{sec:observability} guarantees that the augmented state is locally reconstructable from biomass and substrate measurements alone, so that the UKF can track both $x$ and $\theta$ even during the intervals between DNA measurements.}

{\color{\revised}The unscented transform parameters were selected as $(\alpha,\beta,\kappa)=(0.5,2,0)$. The initial covariance matrix was chosen to reflect moderate uncertainty on the states and a $15\%$ uncertainty on the nominal parameter vector $\theta_{\mathrm{nom}}$, consistently with the Monte Carlo validation setup. Small process-noise covariances were introduced both on the states and on the random-walk parameter dynamics in order to improve robustness against model mismatch and slow biological variability.}
\ifpreprint
Complete tuning parameters and filter covariances are reported in Table~\ref{tab:ukf_tuning}.
\begin{table*}[h]
\centering
\caption{UKF tuning parameters and covariance initialization.}
\label{tab:ukf_tuning}
\renewcommand{\arraystretch}{1.15}
\begin{tabular}{ll}
\hline
\textbf{Quantity} & \textbf{Value} \\
\hline
Initial state std.\ dev.\ & $\sigma_{[b,s,x]}=[0.05,\ 0.05,\ 0.20]$ \\
Initial parameter std.\ dev.\ & $\sigma_{\theta}=0.15\,\theta_{\mathrm{nom}}$, $\theta=[\mu_{\max},K_s,c,Y]^\top$ \\
Process noise std.\ dev.\ & $\sigma_Q=[10^{-2},10^{-2},10^{-3},2{\cdot}10^{-4},2{\cdot}10^{-5},2{\cdot}10^{-5},2{\cdot}10^{-4}]$ \\
Measurement noise std.\ dev.\ & $\sigma_R=[2{\cdot}10^{-2},2{\cdot}10^{-2},10^{-4}]$ \\
UKF scaling parameters & $\alpha=0.5,\ \beta=2,\ \kappa=0$ \\
\hline
\end{tabular}
\end{table*}
\fi

{\color{\revised}The UKF runs at period $T_{\mathrm{UKF}}$, faster than the MPC update period $T_s$.} At each MPC update instant $t_k$, the latest
UKF estimates are used to initialize the prediction model, that is, $\bar{\xi}_{k|k} = \hat{\xi}(t_k)$ and $\theta = \hat{\theta}(t_k)$.

{\color{\revised}The MINLP~\eqref{eq:mpc_cost}--\eqref{eq:mpc_xmax} is then solved with the current parameter estimate $\hat{\theta}(t_k)$ in place of the true $\theta$, and the resulting dilution rate $D_k$ and filtration command $\delta_k$ are applied to the plant. Between consecutive MPC updates, the previously computed input is held constant while the UKF continues to update the state and parameter estimates.}

\subsection{Numerical validation}
\label{sec:numerical_validation}

\begin{figure}
    \centering    \includegraphics[width=0.85\linewidth]{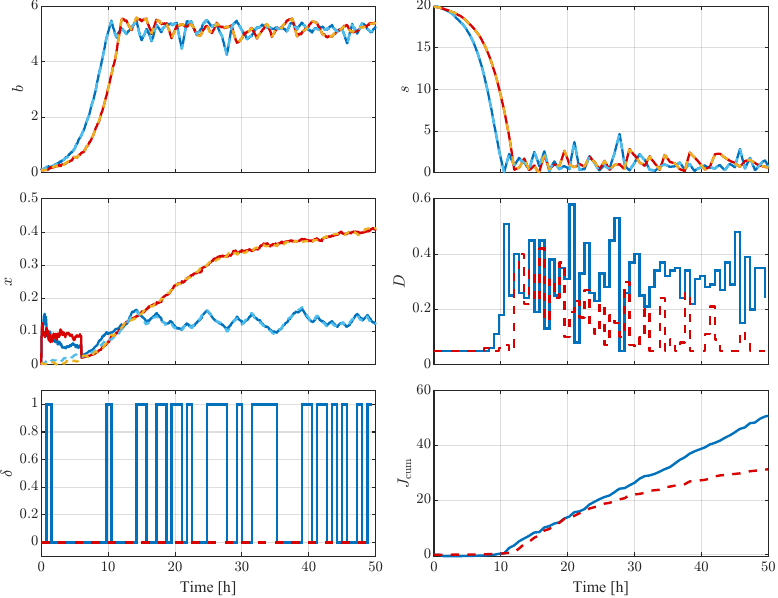}
    \caption{\footnotesize {\color{\revised}Representative closed-loop trajectories comparing the proposed UKF-MPC strategy against the baseline. Blue and red curves denote the active-filtration and no-filtration strategies, respectively. For biomass $b$, substrate $s$, and extracellular DNA concentration $x$, solid lines show UKF estimates and dashed lines show the corresponding true states. The other panels report the dilution rate $D$, the filtration command $\delta$, and the cumulative economic gain $J_{\mathrm{cum}}$.}}
\label{fig:representative_states_inputs_gain}
\end{figure}

{\color{\revised}The proposed UKF-MPC strategy with active electrophoretic filtration was validated against a UKF-MPC baseline with filtration disabled, i.e., $\delta_k = 0$ for all $k$.
The validation was carried out through a paired Monte Carlo study comprising $50$ closed-loop simulations under parametric uncertainty. In each scenario, the kinetic parameters $\theta = [\mu_{\max}, K_s, c, Y]^\top$ were independently sampled uniformly within $\pm 15\%$ of their nominal values reported in Table \ref{tab:parameters}. For each realization, both control strategies were simulated under identical initial conditions, $b(0)=0.1\,\mathrm{g\,L^{-1}}$, $s(0)=20\,\mathrm{g\,L^{-1}}$, $x(0)=0\,\mathrm{ng\,\mu L^{-1}}$, and noise sequences in order to ensure a consistent closed-loop comparison.}
A representative closed-loop trajectory is shown in Fig. \ref{fig:representative_states_inputs_gain}.

\subsubsection{Estimation performance}

{\color{\revised}The UKF estimation accuracy was assessed by computing the RMSE between estimated and true state and parameter trajectories over the closed-loop simulations. The estimation errors remained small over the full horizon. In particular, biomass and substrate RMSEs were below $1.3\times10^{-2}\,\mathrm{g\,L^{-1}}$, while the DNA-state RMSE remained below $2.6\times10^{-2}\,\mathrm{ng\,\mu L^{-1}}$ despite the $6\,\mathrm{h}$ inter-measurement interval. Parameter-estimation errors also remained bounded, with RMSEs below $1.9\times10^{-2}$ for all estimated parameters. The numerical consistency of the filter was monitored at each update step through the augmented covariance matrix $P_{z,k}$. In all Monte Carlo runs, $P_{z,k}$ remained positive definite with a maximum condition number of order $10^{4}$. These results indicate that the UKF provides stable state and parameter estimates along the closed-loop trajectories, consistently with the local observability and identifiability property established in Proposition~\ref{prop:obs}.}
\ifpreprint
Complete estimation performance statistics are reported in Table~\ref{tab:ukf_results}.
\begin{table}[h]
\centering
\caption{UKF estimation performance over the $50$ Monte Carlo
scenarios. Values are mean $\pm$ standard deviation.}
\label{tab:ukf_results}
\scriptsize
\setlength{\tabcolsep}{2.5pt}
\renewcommand{\arraystretch}{1.05}
\begin{tabular}{@{}lcc@{}}
\hline
Metric & With filtration & Baseline \\
\hline
$b$ RMSE [g\,L$^{-1}$]
  & $1.26{\times}10^{-2}\!\pm\!2.43{\times}10^{-4}$
  & $1.26{\times}10^{-2}\!\pm\!1.94{\times}10^{-4}$ \\
$s$ RMSE [g\,L$^{-1}$]
  & $1.28{\times}10^{-2}\!\pm\!3.48{\times}10^{-4}$
  & $1.28{\times}10^{-2}\!\pm\!2.36{\times}10^{-4}$ \\
$x$ RMSE [ng\,$\mu$L$^{-1}$]
  & $2.40{\times}10^{-2}\!\pm\!5.84{\times}10^{-3}$
  & $2.58{\times}10^{-2}\!\pm\!6.25{\times}10^{-3}$ \\
$\mu_{\max}$ RMSE [h$^{-1}$]
  & $1.54{\times}10^{-2}\!\pm\!6.42{\times}10^{-3}$
  & $1.87{\times}10^{-2}\!\pm\!5.68{\times}10^{-3}$ \\
$K_s$ RMSE [g\,L$^{-1}$]
  & $1.77{\times}10^{-3}\!\pm\!8.56{\times}10^{-4}$
  & $1.73{\times}10^{-3}\!\pm\!9.28{\times}10^{-4}$ \\
$c$ RMSE
  & $8.46{\times}10^{-4}\!\pm\!4.02{\times}10^{-4}$
  & $9.93{\times}10^{-4}\!\pm\!5.01{\times}10^{-4}$ \\
$Y$ RMSE
  & $7.70{\times}10^{-3}\!\pm\!3.50{\times}10^{-3}$
  & $9.90{\times}10^{-3}\!\pm\!1.67{\times}10^{-3}$ \\
$\min\,\lambda(P_{z,k})$
  & $1.00{\times}10^{-8}$
  & $1.00{\times}10^{-8}$ \\
$\max\,\kappa(P_{z,k})$
  & $7.63{\times}10^{4}$
  & $1.17{\times}10^{5}$ \\
\hline
\end{tabular}
\end{table}
\fi

\subsubsection{Closed-loop control performance}

{\color{\revised}The two strategies were compared consistently with the validation framework described in Section~\ref{sec:validation_det}. The UKF-MPC with active filtration achieved a statistically significant improvement in terminal cumulative economic return over the baseline, with an average scenario-wise relative increase of $67.02\%\pm14.96\%$ ($p\approx 0$). The final biomass concentrations were statistically indistinguishable between the two strategies, with a relative difference of $-0.13\%\pm6.00\%$ ($p=0.943$), confirming that the performance gain reflects a more efficient closed-loop operating policy rather than a higher terminal biomass. In particular, active DNA removal allowed the controller to sustain a mean dilution rate increased by $108.49\%\pm20.17\%$ ($p\approx 0$). At the same time, the peak extracellular-DNA concentration was reduced  of $58.79\%\pm2.45\%$ ($p\approx 0$). The mechanism underlying this improvement is therefore the ability of the filtration-enabled controller to combine higher dilution rates with active suppression of the inhibitory DNA state, which would otherwise limit the achievable biomass productivity (see Table \ref{tab:mc_statistical_comparison}).}

\subsubsection{Computational tractability}
{\color{\revised}The MPC solution time was recorded at each optimization step over all $50$ scenarios.
The worst-case solution time was $36.69\,\mathrm{s}$, corresponding to $1.36\%$ of $T_s$, confirming real-time implementability at the considered sampling rate.}

\section{Conclusions}
\label{sec:conclusions}

We addressed the feedback control of a recirculating bioreactor in which extracellular DNA accumulates and inhibits microbial growth. {\color{\revised}The problem was formulated as a hybrid economic optimal control problem, with a continuous dilution-rate input and a binary electrophoretic filtration-activation input coordinated over a receding horizon.}

{\color{\revised}In the Monte Carlo simulations, active electrophoretic DNA removal enabled a statistically significant improvement in cumulative economic return of approximately $67\%$ over a filtration-free UKF-MPC baseline, with no statistically significant difference in final biomass concentration ($p=0.943$). The performance gain was attributable to the ability of the filtration-enabled controller to sustain higher dilution rates while actively suppressing the inhibitory DNA state.}

{\color{\revised}Future work will target hardware-in-the-loop validation, in which the proposed control architecture will be tested against real-time measurements from the electrophoretic filtration prototype currently under experimental development.}

\section*{Acknowledgements}
\footnotesize{
Project co-funded by the European Union – Next Generation EU - under the National Recovery and Resilience Plan (NRRP), Mission 4 Component 2, Investment 3.3 – Decree no. 630 (24th April 2024)  of Italian Ministry of University and Research;
Concession Decree no. 1956 del 05th December 2024 adopted by the Italian Ministry of University and Research, CUP D93D24000270003, within the national PhD Programme in Autonomous Systems (XL cycle), and NOSELF s.r.l.
The authors thank Prof.\ S.\ Mazzoleni and Prof.\ F.\ 
Giannino (Dept.\ of Agricultural Sciences, University of 
Naples Federico II) for discussions on inhibitory 
extracellular DNA modelling. AS acknowledges partial PhD 
support from NOSELF s.r.l. AI-assisted tools were used for 
manuscript revision and grammar checking.}

\bibliographystyle{ieeetr}
\bibliography{reference}

\ifpreprint
\appendix
\subsection{Proof of Proposition \ref{prop:obs}}
\normalsize
\label{app:proof_prop1}
%
We apply the nonlinear observability rank condition of Hermann and Krener~\cite{HermannKrener1977}. Consider the deterministic augmented state
\begin{equation}
z = [b,\, s,\, x,\, \mu_{\max},\, K_s,\, c,\, Y]^\top \in \mathbb{R}^7,
\end{equation}
with augmented drift vector field $f_a(z,u)$ obtained from~\eqref{eq:model} by appending trivial parameter dynamics $\dot{\theta}=0$. The measured outputs are $h_1(z)=b$ and $h_2(z)=s$. We show that the observability codistribution generated by the differentials of the outputs and their iterated Lie derivatives along~$f_a$ has full rank in the admissible operating region.

\emph{Step~1 (Zeroth-order).}
The differentials $\mathrm{d}h_1=\mathrm{d}b$ and $\mathrm{d}h_2=\mathrm{d}s$ are linearly independent, yielding rank~$2$.

\emph{Step~2 (First Lie derivative of~$h_1$).}
Since
\begin{equation}
L_{f_a} h_1 = \dot{b} = \mu(s,x)b - Db,
\end{equation}
we have
\begin{equation}\label{eq:dLfh1}
\frac{\partial (L_{f_a} h_1)}{\partial x}
=
-\frac{\mu_{\max}\, s}{(K_s + s)\, x_{\mathrm{crit}}}\, b \neq 0,
\end{equation}
whenever $b>0$ and $s>0$. Hence $\mathrm{d}(L_{f_a} h_1)$ has a nonzero component along~$\mathrm{d}x$, independent of $\{\mathrm{d}b,\mathrm{d}s\}$, and the rank increases to~$3$.

\emph{Step~3 (First Lie derivative of~$h_2$).}
From
\begin{equation}
L_{f_a} h_2
=
\dot{s}
=
-\frac{1}{Y}\mu(s,x)b + D(s_{\mathrm{in}} - s),
\end{equation}
the partial derivative with respect to the yield coefficient is
\begin{equation}\label{eq:dLfh2_Y}
\frac{\partial (L_{f_a} h_2)}{\partial Y}
=
\frac{\mu_{\max}\, s\, (x_{\mathrm{crit}} - x)}
{Y^2\, x_{\mathrm{crit}}\, (K_s + s)}\, b \neq 0,
\end{equation}
under the standing assumptions and $x < x_{\mathrm{crit}}$. Since $Y$ does not appear in $L_{f_a} h_1$, the differential $\mathrm{d}(L_{f_a} h_2)$ introduces an independent component along~$\mathrm{d}Y$, raising the rank to~$4$.

\emph{Step~4 (Second Lie derivatives).}
The parameter~$c$ governs extracellular DNA production and does not appear in the zeroth- or first-order Lie derivatives, since $\dot{b}$ and $\dot{s}$ do not directly depend on~$c$. It enters through the extracellular DNA dynamics $\dot{x} = c\,\mu\,b - \cdots$, so that algebraic manipulation gives
\begin{equation}\label{eq:dLf2h1_c}
\frac{\partial (L_{f_a}^2 h_1)}{\partial c}
=
-\frac{\mu_{\max}^2\, s^2\, (x_{\mathrm{crit}} - x)}
{x_{\mathrm{crit}}^2\, (K_s + s)^2}\, b^2 \neq 0.
\end{equation}
This term is independent of all previously obtained codirections, so $\mathrm{d}(L_{f_a}^2 h_1)$ raises the rank to~$5$. Therefore, the first five rows of $\mathcal{O}(z)$ are analytically proven to be linearly independent.

From this point onward, higher-order Lie derivatives generate lengthy nonlinear expressions involving coupled combinations of states and parameters, making a complete closed-form symbolic proof impractical within the scope of the paper. The observability analysis is therefore partly analytical and partly computational.

To complete the rank verification, the Lie derivatives $L_{f_a}^k h_1$ and $L_{f_a}^k h_2$ were computed symbolically up to order $k=4$ using the \textsc{Matlab} Symbolic Math Toolbox. The resulting $10 \times 7$ observability matrix $\mathcal{O}(z)$ was then evaluated over $10^8$ randomly sampled operating points spanning the admissible parameter and state ranges. In all tested cases, $\mathrm{rank}\,\mathcal{O}(z)=7=\dim(z)$.
The result therefore indicates that, in the admissible operating region, the augmented nonlinear system satisfies the Hermann--Krener local observability rank condition. In particular, direct extracellular DNA measurements are not structurally required for local reconstruction of the hidden DNA state and the uncertain kinetic parameters from biomass and substrate measurements alone.

\fi

\end{document}